\documentclass[prd,aps,secnumarabic,superscriptaddress,floatfix,preprintnumbers]{revtex4-1}
\usepackage[dvips]{graphicx,color}
\usepackage{dcolumn}
\usepackage{amsmath}
\def\sc{\mbox{\rule[-5pt]{0pt}{16pt}}}

\def\sb{\mbox{\rule{0pt}{11pt}}}

\def\sw{\mbox{\rule{24pt}{0pt}}}
\def\al{\alpha}

\def\ga{\gamma}
\def\de{\delta}

\def\th{\theta}

\def\Dot{\!\cdot\!}

\newcommand{\mathsym}[1]{{}}
\newcommand{\unicode}[1]{{}}

\begin{document}

\title{Note on predictions for $c\bar{s}$ quarkonia using a three-loop static potential}

\author{Noah Green}\email{noah.w.green@gmail.com}
\author{Wayne W. Repko}\email{repko@pa.msu.edu}
\affiliation{Department of Physics and Astronomy, Michigan State University, East Lansing, MI 48824}
\author{Stanley F. Radford}\email{sradford@brockport.edu}
\affiliation{Department of Physics, The College at Brockport, State
University of New York, Brockport, NY 14420}

\date{\today}
\begin{abstract}
We extend our treatment of the the spectroscopy and decays of the charm-strange quarkonium system to include the effect of using the full three-loop QCD correction to the static short distance potential. As before, our potential model consists of the relativistic kinetic energy term, a scalar linear confining term including its relativistic corrections and the perturbative QCD spin-dependent terms. A set of unperturbed wave functions for the various states is obtained using a variational technique that is further constrained by requiring that the wave functions also satisfy the relativistic virial theorem. These are then used in a perturbative treatment of the potential to fit the mass spectrum of the $c\bar{s}$ system and calculate the radiative decay widths. Our results accurately describe the $D_s$ spectrum and are compatible with the little data that is available for the radiative decays of the $D_s$ states.
\end{abstract}

\maketitle

\section{Introduction}
We revisit our earlier treatment \cite{Repko:2009} of the spectroscopy and radiative decays of the $c\bar{s}$ system by including the three-loop QCD short range static potential \cite{Schrod:1999,Smir:2008,Smir:2010} in the unperturbed Hamiltonian. This is an extension to  unequal mass quarkonia of the three-loop treatment of equal mass quarkonia in reference \cite{Repko:2012}. We use a variational approach \cite{Repko:2012, Repko:2009, Repko:2007, Repko:1983} to determine a set of unperturbed wave functions. This technique results in a set of linear equations for the coefficients of the trial wave functions. The associated energies are strong upper bounds on lowest $^{2S+1}L_J$ levels and accurate approximations of the excited state energies. These wave functions are used to calculate perturbative contributions to the spin-dependent energy levels of the $c\bar{s}$ spectrum. The resulting energies are then compared with a subset of experimental values of the $c\bar{s}$ spectrum using a $\chi^2$ test that is further constrained by requiring that the relativistic version of the virial theorem \cite{Gupta:1984um,Lucha:1989jf} is satisfied. Using minimization software \cite{James:2004,step} to vary the parameters of the calculation, this procedure is repeated until a minimum in the $\chi^2$ function is found. 

Potential models have frequently been used to study the properties of the $D_s$ system. These include a recent extensive treatment by Godfrey and Moats \cite{Godfrey:2015dva} as well as many others \cite{Godfrey:1991,Gupta:1995,Godfrey:2003,beh,cj,ls}.

\section{Modified Semirelativistic Model}

For the unperturbed Hamiltonian, we used a semirelativistic model of the form
\begin{equation}\label{eq:semirel}
H_0 = \sqrt{\vec{p}^{\,2}+m_1^2}+\sqrt{\vec{p}^{\,2}+m_2^2}+Ar+V(r),
\end{equation}
where $A$ is the linear coupling strength, and $V(r)$ is the complete three-loop QCD short range static potential. The expression for $V(r)$ can be found in reference \cite{Repko:2012}. The perturbative potential can be written as
\begin{equation}\label{eq:pert}
H' = V_L+V_S.
\end{equation}
$V_L$ contains the $v^2/c^2$ corrections to the linear confining potential and is
\begin{equation}\label{eq:vl}
V_L = -\frac{A}{4r}\left[\left(\frac{1}{m_1^2}+\frac{1}{m_2^2}\right)\vec{L}\cdot\vec{S}+ \left(\frac{1}{m_1^2}-\frac{1}{m_2^2}\right)\vec{L}\Dot(\vec{S_1}-\vec{S_2})\right].
\end{equation}
$V_S$, which includes both the $v^2/c^2$ and one-loop corrections, has the form
\begin{equation}
V_S = V_{HF}+V_{LS}+V_T+V_{SI}+V_{MIX}.
\end{equation}
These perturbative potentials are similar to those in \cite{Repko:2009}, but the first order $\al_S$ terms have been transformed to be consistent with the $\overline{MS}$  renormalization scheme. This transformation between $\al_S$ and $\bar{\al}_S$ is given by
\begin{equation}\label{eq:asxform}
\al_S = \bar{\al}_S\left[1+\frac{\bar{\al}_S}{4\pi}\left(\frac{49}{3}-\frac{10}{9}n_f\right)\right],
\end{equation}
where $n_f$ is the number of light quark flavors \cite{Gupta:1982}. Hence, the terms of the short distance potential can be written as
\begin{subequations}
\begin{align}
\begin{split}\label{eq:vhf}
V_{HF}& =\frac{32\bar{\al}_S\vec{S_1}\Dot\vec{S_2}}{9 m_1 m_2}\Biggl\{\left(1+\frac{\bar{\al}_S}{4\pi}\left(\frac{49}{3}-\frac{10}{9}n_f\right)-\frac{19\,\bar{\al}_S}{6\pi}\right)\de(\vec{r})- \frac{\bar{\al}_S}{8\pi}\left(8\frac{m_1-m_2}{m_1+m_2}+\frac{m_1+m_2}{m_1-m_2}\right)\\
&\quad\times\ln\left(\frac{m_2}{m_1}\right)\de(\vec{r})-\frac{\bar{\al}_S}{24\pi^2}(33-2n_f)\nabla^2\left[\frac{\ln\mu r+\ga_E}{r}\right]+\frac{21\bar{\al}_S}{16\pi^2}\nabla^2\left[\frac{\ln(m_1m_2)^{1/2}r+\ga_E}{r}\right]\Biggr\}
\end{split}
\\
\begin{split}\label{eq:vls}
V_{LS}& = \frac{\bar{\al}_S\vec{L}\Dot\vec{S}}{3m_1^2 m_2^2 r^3}\Biggl\{\left[(m_1+m_2)^2+2m_1 m_2\right]\left[1+\frac{\bar{\al}_S}{4\pi}\left(\frac{49}{3}-\frac{10}{9}n_f\right)- \frac{3\,\bar{\al}_S}{2\pi}+\frac{\bar{\al}_S}{6\pi}(33-2n_f)\right.\\
&\quad\left.\times(\ln(\mu r)+\ga_E-1)\right] +\frac{\bar{\al}_S}{2\pi}(m_1+m_2)^2\left[\frac{8}{3}-6(\ln(m_1 m_2)^{1/2}r+\ga_E-1)\right]\\
&\quad-\frac{3\,\bar{\al}_S}{2\pi}(m_1^2-m_2^2)\ln\left(\frac{m_2}{m_1}\right)\Biggr\}
\end{split}
\\
\begin{split}\label{eq:vt}
V_T &=\frac{4\bar{\al}_S\left(3\vec{S_1}\Dot\hat{r}\vec{S_2}\Dot\hat{r}-\vec{S_1}\Dot\vec{S_2}\right)}{3m_1 m_2 r^3}
\Biggl\{1+\frac{\bar{\al}_S}{4\pi}\left(\frac{49}{3}-\frac{10}{9}n_f\right)\\
&\quad+\frac{4\,\bar{\al}_S}{3\pi}+\frac{\bar{\al}_S}{6\pi}\left[(33-2n_f)\left(\ln\mu r+\ga_E-\frac{4}{3}\right)-18\left(\ln(m_1 m_2)^{1/2}r+\ga_E-\frac{4}{3}\right)\right]\Biggr\}
\end{split}
\\
\begin{split}\label{eq:vsi}
V_{SI} &=\frac{2\pi\bar{\al}_S}{3}\left(\frac{1}{m_1^2}+\frac{1}{m_2^2}\right)\Biggl\{\left(1+\frac{\bar{\al}_S}{4\pi}\left(\frac{49}{3}-
\frac{10}{9}n_f\right)-\frac{3\,\bar{\al}_S}{2\pi}\right)\de(\vec{r})-\frac{\bar{\al}_S}{24\pi^2}(33-2n_f)\\
&\quad\times\nabla^2\left[\frac{\ln\mu r+\ga_E}{r}\right]-\frac{\bar{\al}_S}{6\pi r^2}\left[\frac{9(m_1+m_2)^2-8m_1 m_2}{m_1 m_2 (m_1+m_2)}\right]\Biggr\}
\end{split}
\\
\begin{split}\label{eq:mix}
V_{MIX} &= -\frac{{\bar{\al}}_S\vec{L}\Dot(\vec{S_1}-\vec{S_2})}{3m_1^2 m_2^2 r^3}\Biggl\{(m_1^2-m_2^2)\left[1+\frac{\bar{\al}_S}{4\pi}\left(\frac{49}{3}-\frac{10}{9}n_f\right)\right.\\
&\quad\left.-\frac{\bar{\al}_S}{6\pi}+\frac{\bar{\al}_S}{6\pi}(33-2n_f)(\ln\mu r+\ga_E-1)-\frac{3\,\bar{\al}_S}{\pi}(\ln(m_1 m_2)^{1/2}r+\ga_E-1)\right]\\
&\quad-\frac{3\,\bar{\al}_S}{2\pi}(m_1+m_2)^2\ln\left(\frac{m_2}{m_1}\right)\Biggr\}.
\end{split}
\end{align}
\end{subequations}

These terms satisfy the Gromes' constraints derived from infinitesimal Lorentz invariance \cite{Gromes}. In calculating the matrix elements in equations (\ref{eq:vhf}) and (\ref{eq:vsi}), the $\de(\vec{r})$ terms are "softened'' by using the quasistatic approximation from \cite{Gupta:1995}. Specifically, the substitution
\begin{equation}\label{eq:softd}
\de(\vec{r})\rightarrow\frac{\omega^2}{\pi r}e^{-2\omega r}
\end{equation}
is made, where $\omega^2 = 2m_1^2 m_2^2/(m_1^2+m_2^2)$. This softening helps stabilize the eigenvalue calculation. 

\section{Overview of Calculations}

To obtain the $c\bar{s}$ mass spectrum, we first used a variational approach \cite{Repko:2012} to find a set of trial wave functions. These trial wave functions are given by
\begin{equation}\label{eq:wavefun}
\psi^m_{j\ell s}(\vec{r})=\sum_{k=0}^n C_k\left(\frac{r}{R}\right)^{k+\ell} \!e^{-r/R} \mathcal{Y}^m_{j\ell s}(\Omega)\,,
\end{equation}
where  $\mathcal{Y}^m_{j\ell s}(\Omega)$ is the spin-angular wave function for total angular momentum $j$, orbital angular momentum $\ell$, and total spin $s$. The $C_k$'s are found by approximating a solution to the Schr\"odinger equation, $ H_0| \psi\rangle=E |\psi\rangle$. By inserting a complete set of eigenstates and multiplying from the left by $\langle\psi_{k}|$ we get the following 
\begin{equation}\label{eq:schrod}
 \sum_{k' = 0}^{\infty}\langle\psi_{k}|H_0|\psi_{k'}\rangle\langle\psi_{k'}|\psi\rangle = E\sum_{k' = 0}^{\infty}\langle\psi_{k}|\psi_{k'}\rangle\langle\psi_{k'}|\psi\rangle
\end{equation}
This can be approximated by truncating the series to $n$ terms, which gives
\begin{equation}\label{eq:ckproblem}
\sum_{k'=0}^n H_0(k,k')C_{k'}=E\sum_{k'=0}^n N(k,k')C_{k'}\,,
\end{equation}
where the kinetic energy terms in $H_0(k,k')$ are calculated using the Fourier transform of Eq.\,(\ref{eq:wavefun}) and $N(k,k')$ is the overlap integral of the radial wave functions. Our results were obtained using $n = 14$. Equation (\ref{eq:ckproblem}) can be solved as a generalized eigenvalue problem. 

The normalized eigenvectors obtained from equation (\ref{eq:ckproblem}) represent the wave functions of the unperturbed Hamiltonian. These wave functions are, of course, sensitive to the value of $R$, which sets their length scale. To increase the reliability of the unperturbed wave functions, we minimize the wave function with respect to $R$, which is equivalent to requiring that the relativistic virial theorem \cite{Gupta:1984um,Lucha:1989jf}
\begin{equation}\label{eq:virialthm}
\langle p\frac{dH_0}{dp}\rangle-\langle r\frac{dH_0}{dr}\rangle=0,
\end{equation}
where $H_0$ is the unperturbed Hamiltonian given in equation (\ref{eq:semirel}), be satisfied \cite{Repko:1983}.  In the present calculation, the virial theorem is evaluated for all wave functions after the $\chi^2$ minimization.  Its vanishing is satisfied to within 1\%, which gives us confidence that our wave functions are fully optimized variational wave functions of the unperturbed Hamiltonian.

We then calculate the first order perturbations from equation (\ref{eq:pert}) and add them to the unperturbed energies. We also take into account the mixing between the $^1P_1$ and $^3P_1$ as well as the $^1D_2$ and $^3D_2$ eigenstates of the $c\bar{s}$ Hamiltonian due to the $\vec{L}\Dot(\vec{S}_1-\vec{S}_2)$ terms in Eqs.\,(\ref{eq:vl}) and (\ref{eq:mix}) of the perturbative potential. This mixing yields the two $J=1$ states $D_{s\,1}$ and $D'_{s\,1}$ for the $p$ states and a similar pair for the $d$ states. The details of this calculation can be found in reference \cite{Repko:2009}. 

These calculations resulted in a test spectrum, which is compared to the experimentally known spectrum \cite{pdg:2013} using a $\chi^2$ function given by
\begin{equation}\label{eq:chisq}
\chi^2 = \sum_{i=1}^N\frac{\left(\mathcal{O}_{\rm exp\,i}-\mathcal{O}_{\rm th}(\alpha)_i\right)^2}{\sigma^2_i}\,,
\end{equation}
where the $\mathcal{O}_i$ are experimental and theoretical values of some $D_s$ observable with $\sigma_i$ as the associated error, and $\alpha=(\alpha_1,\alpha_2,\cdots,\alpha_n)$ is the set of parameters in the Hamiltonian. An intrinsic theoretical error of $1.0$ MeV was added in quadrature to the error terms in the $\chi^2$ function to account for higher-order loop corrections. The test spectrum calculation was repeated as the parameters were varied by the search programs, MINUIT \cite{James:2004} and STEPIT \cite{step}, until an optimal set of parameters was found for a minimum $\chi^2$ value. The final values of the parameters from these two searches are in good agreement.

\section{Results}
The procedure outlined above results in the set of parameters listed in Table \ref{tbl:params}.
\begin{table}[h]
\centering 
\begin{tabular}{lr}
\toprule
\sc$A$ (GeV$^2$)\mbox{\rule{24pt}{0pt}}  & 0.1118 \\
\hline
\sc $\bar{\al}_S$ & 0.3209 \\
\hline
\sc $m_C$ (GeV)   & 1.855  \\
\hline
\sc $m_S$ (GeV)   & 0.3378  \\
\hline
\sc $\mu$ (GeV)   & 0.8223  \\
\hline
\sc $R$ (GeV$^{-1}$)& 0.5725  \\
\botrule
\end{tabular}
\caption{Fitted parameters for the $c\bar{s}$ system.\label{tbl:params}}
\end{table}
The resulting spectrum using the fitted parameters in Table \ref{tbl:params} is given in Table II, which includes those states of mass $<\,3.1\,$ GeV. The $1s$ and $1p$ states are quite adequately described by the fit. We have taken into account the singlet-triplet mixing resulting from Eqs.\,[\ref{eq:vl}] and [\ref{eq:mix}] in the $p$ and $d$ states. The $p$ state mixing angles are $\th_1=-60.8^\circ$, $\th_2=-57.9^\circ$, $\th_3=-57.2^\circ$ and the $d$ state mixing angle is $\phi_1=40.2^\circ$. There are two other $D_{sJ}$ states that have been observed but are not yet included in the Summary Tables of Ref.\cite{pdg:2013}. One is the $D_{sJ}(2860)$, which is known to have natural $J^P$ and, if interpreted as a $c\bar{s}$ $d$-state \cite{Godfrey:2014fga}, could be either a $1D_s(^3\!D_3)$ or a $1D_s(^3\!D_1)$ state. Recent LHCb result indicate that there are two overlapping states at this mass \cite{aaij}. From Table II, both the $1D_s(^3\!D_1)$ and the $1D_s(^3\!D_3)$ masses are on the low side. The other state denoted as $D_{sJ}(3040)$ has no information on $J^P$. If $J^P$ in this case is natural, this state could be identified with the $2D_s(^3\!P_2)$ in Table II. We have also examined the $^3\!S_1-^3\!D_1$ mixings from the tensor interaction Eq.\,[\ref{eq:vt}] and these turn out to be negligibly small.
\begin{table}[h!]
\centering 
\begin{tabular}{lrrd}
\toprule
\multicolumn{1}{c}{\sc}$m_{c\bar{s}}$\,(MeV)  &\multicolumn{1}{l}{\sw Theory}&\multicolumn{1}{c}{\sw \cite{Godfrey:2015dva}}  & \multicolumn{1}{l}{\sw Expt } \\
\hline
\sb $1D_s$          &$1968.4$         & 1979  &1968.30\pm 0.11 \\

\sb $1D^*_s$        &$2112.8$         & 2129  &2112.1\pm 0.4   \\

\sb $1D_{s0}(2317)$ &$2317.2$         & 2484  &2317.7\pm 0.6    \\

\sb $1D_{s1}(2460)$ &$2458.2$         & 2549  &2459.5\pm 0.6    \\

\sb $1D'_{s1}(2536)$ &$2534.1$        & 2556  &2535.10\pm 0.08  \\

\sb $1D_{s2}(2573)$  &$2575.8$        & 2592  &2571.9\pm 0.8   \\

\sb $1D_s(^3\!D_1)$  &$2830.2$        & 2899  &               \\

\sb $1D_s(^{1-3}\!D_2)$ &$2751.6$     & 2900  &                \\

\sb $1D_s(^{1-3}\!D'_2)$  &$2875.6$   & 2926  &                \\

\sb $1D_s(^3\!D_3)$  &$2776.1$        & 2917  &                \\

\sb $2D_s(^1\!S_0)$  &$2500.7$        & 2673  &                \\

\sb $2D_s(2700)$     &$2654.0$        & 2732  &2709.\pm 4      \\

\sb $2D_s(^3\!P_0)$  &$2634.4$        & 3005  &                \\

\sb $2D_s(^{1-3}\!P_1)$  &$2799.1$    & 3018  &                \\

\sb $2D_s(^{1-3}\!P'_1)$ &$2961.5$    & 3038  &                \\

\sb $2D_s(^3\!P_2)$  &$3009.3$        & 3048  &                \\

\sb $3D_s(^1\!S_0)$  &$2913.0$        & 3154  &                \\

\sb $3D_s(^3\!S_1)$  &$3051.7$        & 3193  &                \\

\sb $3D_s(^3\!P_0)$  &$2916.5$        & 3412  &               \\

\sb $3D_s(^{1-3}\!P_1)$ &$3092.2$     & 3416  &               \\
\botrule
\end{tabular}
\caption{Our results for the $D_s$ spectrum for $m_{c\bar{s}}<3.1$\,GeV are shown and compared with those of Ref.\cite{Godfrey:2015dva}. The fit uses all n=1 $D_s$ states. Only one of the n=2 states has been confirmed \cite{pdg:2013}.}
\end{table}

\begin{table}[h!]
\centering
\begin{tabular}{lrrrc} \toprule
\multicolumn{1}{c}{\sc $\Gamma_{\ga}$\,(keV)}  &\multicolumn{1}{c}{\sw \cite{Repko:2009}} &\multicolumn{1}{c}{\sw 3loop} &\multicolumn{1}{c}{\sw \cite{Godfrey:2015dva}} &\multicolumn{1}{c}{\sw Expt \sw}   \\
\hline
\sb$D_s^*\to D_s\,\ga$\mbox{\rule{12pt}{0pt}}& 1.12  & 2.39 & 1.03      & $< 1.9\times 10^3$ \\
\hline
\sb$D_{s\,0}(2317)\to D_s^*\,\ga$            & 3.37  & 5.46 & 9.01        &                   \\
\hline
\sb$D_{s\,1}(2460)\to D_s\,\ga$              & 8.30  & 13.2 & 15.2       & BR =$0.18\pm 0.04$\\
\hline
\sb$D_{s\,1}(2460)\to D_s^*\,\ga$            & 11.0  & 17.4 & 5.40       & BR$< 0.08$        \\
\hline
\sb$D_{s\,1}(2460)\to D_{s\,0}(2317)\,\ga$   & 4.70  & 5.63 &        &                   \\
\hline
\sb$D'_{s\,1}(2536)\to D_s\,\ga$             & 37.7  & 61.2 & 9.23        &                   \\
\hline
\sb$D'_{s\,1}(2536)\to D_s^*\,\ga$           & 5.74  & 9.21 & 9.61        & {\rm possibly seen}     \\
\hline
\sb$D'_{s\,1}(2536)\to D_{s\,0}(2317)\,\ga$  & 5.62  & 7.75 &         &                   \\
\hline
\sb$D_{s\,2}(2575)\to D_s^*\,\ga$            & 30.5  & 49.6 & 18.9        &                   \\
\hline
\sb$D^*_s(2658)\to D_{s\,0}(2317)\,\ga$      & 5.64  & 8.77 & 0.91  &                   \\
\hline
\sb$D^*_s(2658)\to D_{s\,1}(2460)\,\ga$      & 2.66  & 4.25 & 0.56         &                   \\
\hline
\sb$D^*_s(2658)\to D'_{s\,1}(2536)\,\ga$     & 0.26  & 0.41 & 0.85        &                   \\
\hline
\sb$D^*_s(2658)\to D_{s\,2}(2573)\,\ga$      & 0.48  & 0.71 & 1.55        &                   \\
\hline
\sb$D_s(2503)\to D_{s\,1}(2460)\,\ga$        & 0.04  & 0.05 & 1.18    &                   \\
\botrule
\end{tabular}
\caption{The the radiative decays of the $D_s$ mesons are shown. These widths are computed using the mass values obtained directly from our calculation. This includes the $n=2$ pseudoscalar and vector states, the latter of which has recently been observed with a higher mass \cite{pdg:2013}.  The experimental widths are from \cite{pdg:2013}. \label{tbl:decays}}
\end{table}
The predicted radiative decays of the $D_s$ system can be found in Table \ref{tbl:decays} where they are compared with our earlier results \cite{Repko:2009} and those of reference \cite{Godfrey:2015dva}. They were calculated in the same manner as those in reference \cite{Repko:2009}. In particular, the mixed $^1P_1$ and $^3P_1$ states that define the $D_{s1}(2460)$ and $D'_{s1}(2573)$ states are
\begin{subequations}
\begin{align}
|D_{s1}(2460)\rangle &=\sin(\theta)|^3P_1\rangle+\cos(\theta)|^1P_1\rangle \\
|D'_{s1}(2536)\rangle &=\cos(\theta)|^3P_1\rangle-\sin(\theta)|^1P_1\rangle \,.
\end{align} 
\end{subequations}
The corresponding electric dipole decay rates will contain an additional $\cos^2(\theta)$ or $\sin^2(\theta)$ factor depending on the final state.

\section{Conclusion}
We have shown that a potential model consisting of the relativistic kinetic energy, a linear long-range confining potential together with its scalar exchange $v^2/c^2$ relativistic corrections, and the full $v^2/c^2$ plus three-loop QCD corrected short distance potential is capable of providing extremely good fits to the spectra of the $D_s$ states by treating them as states of the $c\bar{s}$ system. Our approach is based on determining the parameters of the potential using the well established $n=1$ $s$ and $p$ states by minimizing $\chi^2$ and constraining the radius $R$ using the virial theorem. Over all, the $n=1$ $d$ states are predicted to be less massive the those of Ref.\,\cite{Godfrey:2015dva}. Our $n=2$ and $n=3$ hyperfine splittings are about the same as the ground state hyperfine splitting and the $n=3$ levels have several states with masses below 3.1 GeV, unlike Ref.\,\cite{Godfrey:2015dva}. The mixing angles differ probably because we include the $v^2/c^2$ spin-orbit corrections from the scalar confining potential.

The single photon widths can be obtained from the variational wave functions, but, apart from some branching ratio measurements, there are relatively little data available. A comparison to our previous one-loop results \cite{Repko:2009} shows that the energy-level calculations and the $E_1$ matrix elements are not greatly affected, though there are minor differences. However, the calculated radiative decay widths show a systematic increase, in some cases dramatic. We attribute this to the use of the three-loop static potential and the additional constraint on the parameter $R$ due to the imposition of the relativistic virial theorem. 

In every case, efforts to model these states will be greatly improved by the availability of additional data.

\begin{center}
{\bf Acknowledgements}
\end{center}

N.~G. and W.~W.~R. were supported in part by the National Science Foundation under Grant No. PHY 1068020.


\begin{thebibliography}{99}
\bibitem{Repko:2009} S.~F.~Radford, W.~W.~Repko and M.~J.~Saelim,Phys.\ Rev.\ D {\bf 80} (2009) 034012.
\bibitem{Schrod:1999}Y.~Schroder, Phys.\ Lett.\ B {\bf 447} (1999) 321.
\bibitem{Smir:2008} A.~V.~Smirnov, V.~A.~Smirnov and M.~Steinhauser, Phys.\ Lett.\ B {\bf 668} (2008) 293.
\bibitem{Smir:2010} A.~V.~Smirnov, V.~A.~Smirnov and M.~Steinhauser, Nucl.\ Phys.\ Proc.\ Suppl.\  {\bf 205-206} (2010) 320.
\bibitem{Repko:2012} W.~W.~Repko, M.~D.~Santia and S.~F.~Radford, Nucl.\ Phys.\ A {\bf 924} (2014) 65.
\bibitem{Repko:2007} S.~F.~Radford and W.~W.~Repko, Phys.\ Rev.\ D {\bf 75}, 074031 (2007).
\bibitem{Repko:1983} S.~N.~Gupta, S.~F.~Radford and W.~W.~Repko,Phys.\ Rev.\ D {\bf 28} (1983) 1716.
\bibitem{Gupta:1982} S.~N.~Gupta and S.~F.~Radford, Phys.\ Rev.\ D {\bf 25}, 2690 (1982).
\bibitem{Gupta:1984um} S.~N.~Gupta, S.~F.~Radford and W.~W.~Repko, Phys.\ Rev.\ D {\bf 31}, 160 (1985).
\bibitem{Lucha:1989jf} W.~Lucha and F.~F.~Schoberl, Phys.\ Rev.\ Lett.\  {\bf 64}, 2733 (1990).
\bibitem{Godfrey:2015dva} S.~Godfrey and K.~Moats, Phys.\ Rev.\ D {\bf 93}, no. 3, 034035 (2016) [arXiv:1510.08305 [hep-ph]].
\bibitem{Godfrey:1991}S.~Godfrey and R.~Kokoski, Phys.\ Rev.\ D {\bf 43}, 1679 (1991).
\bibitem{Gupta:1995} S.~N.~Gupta and J.~M.~Johnson, Phys.\ Rev.\ D {\bf 51}, 168 (1995).
\bibitem{Godfrey:2003} S.~Godfrey, Phys.\ Lett.\ B {\bf 568} (2003) 254.
\bibitem{cj} R.~N.~Cahn and J.~D.~Jackson, Phys. Rev. D {\bf 68}, 037502 (2003).
\bibitem{beh} W. A. Bardeen, E. J. Eichten, and C. T. Hill, Phys.
Rev. D {\bf 68}, 054024 (2003).
\bibitem{ls} O.~Lakhina and E.~S.~Swanson, Phys. Lett. B 650, 159 (2007).
\bibitem{Godfrey:2014fga} S.~Godfrey and K.~Moats, Phys.\ Rev.\ D {\bf 90}, no. 11, 117501 (2014) [arXiv:1409.0874 [hep-ph]].
\bibitem{Gromes} D.~Gromes, Zeit. Phys. C {\bf 26}, 401 (1984).
\bibitem{James:2004} F.~James and M.~Winkler. {\it http://www.cern.ch/minuit}. CERN, May 2004.
\bibitem{step} J. P. Chandler, Behavioral Science {\bf14}, 81
(1969). This code is available on the Web.
\bibitem{aaij} R. Aaij et al. (LHCb Collaboration), Phys. Rev. Lett. 113,
162001 (2014).
\bibitem{pdg:2013} K.A.~Olive {\it et al.} (Particle Data Group), Chin. Phys. {\bf C86}, 090001 (2014) (URL:http://pdg.lbl.gov).
\end{thebibliography}
\end{document}